\documentclass[prd,amsmath,amssymb,superscriptaddress,preprintnumbers,twocolumn,10pt]{revtex4-1}%superscriptaddress,showpacs,

\usepackage{graphicx}
\usepackage{dcolumn}
\usepackage{bm}
\usepackage{amssymb}
\usepackage{latexsym}
\usepackage{booktabs}
\usepackage{amsmath}
\usepackage{multirow}
\usepackage{url}
\usepackage{footnote}
\usepackage{float}
\usepackage[colorlinks=true, linkcolor=red, citecolor=blue]{hyperref}

\usepackage[normalem]{ulem}
\usepackage{color}
\usepackage{array}
\usepackage{enumerate}
\usepackage{tabularx}   % 智能列宽调整
\usepackage{ragged2e}   % 对齐控制
\usepackage{booktabs}   % 专业表格线 (可选)
\usepackage{tabularx}
\usepackage{booktabs}
\usepackage{amsmath}

\newcommand{\mnras}{Mon. Not. R. Astron. Soc.}
\newcommand{\jcap}{J. Cosmol. Astropart. P.}
\newcommand{\aap}{Astron. Astrophys.}
\newcommand{\apjl}{Astrophys. J. Lett.}

\newcommand{\pasp}{Publications of the Astronomical Society of the Pacific}

\begin{document}

\title{Parameter inference of millilensed gravitational waves using neural spline flows}

\author{Zheng Qin}\thanks{These authors contributed equally to this work.}
%\affiliation{Key Laboratory of Cosmology and Astrophysics (Liaoning) \& College of Sciences, Northeastern University, Shenyang 110819, China}
\affiliation{Key Laboratory of Cosmology and Astrophysics (Liaoning), College of Sciences, Northeastern University, Shenyang 110819, China}

\author{Tian-Yang Sun}\thanks{These authors contributed equally to this work.}
\affiliation{Key Laboratory of Cosmology and Astrophysics (Liaoning), College of Sciences, Northeastern University, Shenyang 110819, China}

\author{Bo-Yuan Li}
\affiliation{Key Laboratory of Cosmology and Astrophysics (Liaoning), College of Sciences, Northeastern University, Shenyang 110819, China}

\author{Jing-Fei Zhang}
\affiliation{Key Laboratory of Cosmology and Astrophysics (Liaoning), College of Sciences, Northeastern University, Shenyang 110819, China}

\author{Xiao Guo}\thanks{Corresponding author; guoxiao17@mails.ucas.ac.cn}
\affiliation{Institute for Gravitational Wave Astronomy, Henan Academy of Sciences, Zhengzhou 450046, Henan, China}
\affiliation{School of Fundamental Physics and Mathematical Sciences, Hangzhou Institute for Advanced Study, University of Chinese Academy of Sciences, No.1 Xiangshan Branch, Hangzhou 310024, China}

\author{Xin Zhang}\thanks{Corresponding author; zhangxin@neu.edu.cn}
%\author{Xin Zhang\footnote{Corresponding author}}
%\email{zhangxin@mail.neu.edu.cn}
\affiliation{Key Laboratory of Cosmology and Astrophysics (Liaoning), College of Sciences, Northeastern University, Shenyang 110819, China}

\affiliation{National Frontiers Science Center for Industrial Intelligence and Systems Optimization, Northeastern University, Shenyang 110819, China}

\affiliation{Key Laboratory of Data Analytics and Optimization for Smart Industry (Ministry of Education), Northeastern University, Shenyang 110819, China}

\begin{abstract}
%当引力波（GWs）的传播路径附近出现一定质量的天体时，将发生引力波(GWs)的引力透镜效应，使得波形发生依赖于透镜模型的调制。这种效应为宇宙学和天体物理学的研究提供了一种有力的工具。由于增加了透镜体参数，以及透镜模型的不确定性，使用传统方法对透镜化GW事件进行参数推断时极为耗时，因此需要更加高效的参数推断方法。在本文中，我们探索了使用神经样条归一化流（NSFs）来对微透镜化GWs进行后验推断。我们的研究表明，与像BILBY这样的基于贝叶斯推断的传统方法相比，我们搭建的NSF网络不仅能够在主要参数上实现不输于传统方法的推断准确度，并且平均可将推断速度从约3天缩短至0.8秒。同时，该网络对于GW源的自旋参数有较好的泛化性。它有望成为未来低延迟搜索透镜信号的有力工具。
When gravitational waves (GWs) propagate near massive objects, they undergo gravitational lensing that imprints lens model dependent modulations on the waveform. 
This effect provides a powerful tool for cosmological and astrophysical studies. 
Due to the added parameters of lenses and the uncertainty of lens models, parameter inference for lensed GW events using traditional methods is extremely time-consuming, thus requiring more efficient parameter inference methods.
In this work, we explore the use of neural spline flows (NSFs) for posterior inference of millilensed GWs, and successfully apply NSFs to the inference of 11-dimensional lens parameters. 
Our results demonstrate that compared with traditional methods like {\tt Bilby dynesty} that rely on Bayesian inference, the NSF network we built not only achieves inference accuracy comparable to traditional methods for most parameters, but also can reduce the inference time from approximately 3 days to 0.8 s on average. 
Additionally, the network exhibits strong generalization for the spin parameters of GW sources. 
It is anticipated to become a powerful tool for future low-latency searches for lensed GW signals.

\end{abstract}
%\pacs{95.36.+x, 98.80.Es, 98.80.-k}

\maketitle
\section{Introduction}
%1.自2015年首次直接探测到引力波以来，LIGO-Virgo-KAGRA (LVK)引力波探测网络的前三个观测期已经报道了93起GW事件，这使得我们可以进行各种新的广义相对论检验[]和宇宙学测试[7,8]。随着对LVK网络第四观测期观测到数据的进一步分析，以及未来各种高性能引力波探测器的兴建（如激光干涉仪空间天线(LISA)[13]、太极[ ]、天琴[ ]、爱因斯坦望远镜(ET)[11]、宇宙探测器(CE) [10]），探测到的GW事件的数量将急剧增加，这将为宇宙学、天文学、天体物理学等领域的研究提供了一种全新的有力工具。而在这些引力波信号中，十分值得被关注的是受透镜效应影响的GWs。引力波的透镜效应是由存在于探测器和波源之间的天体造成的，该效应会导致GWs出现明显的调制【】。
Since the first direct detection of gravitational waves (GWs) in 2015, LIGO-Virgo-KAGRA (LVK) network has reported $93$ GW events in its first three observing runs, which enables us to conduct various novel tests of general relativity \cite{2016PhRvL.116v1101A,2018PhRvL.121l9902A,2019PhRvD.100j4036A,2019PhRvL.123a1102A,2021PhRvD.103l2002A} and cosmological research \cite{2021testsgeneralrelativitygwtc3,Abbott_2023}. 
With the further analysis of the data observed in the fourth observing run of the LVK network and the construction of various high-performance gravitational wave detectors in the future (such as the Laser Interferometer Space Antenna (LISA) \cite{Barausse_2020LISA}, Taiji \cite{Taiji}, TianQin \cite{fundamentalphysicscosmologytianqin}, Einstein Telescope (ET) \cite{Maggiore_2020ET}, and Cosmic Explorer (CE) \cite{evans2021horizonstudycosmicexplorerCE}), the number of detected GW events will increase sharply, providing a brand-new powerful tool for research in fields such as fundamental physics, cosmology, and astrophysics.
Among these GW signals, it is particularly worthy of attention those signals affected by the lensing effect. 
The lensing effect of GWs is caused by celestial bodies located between detectors and sources, which will result in a significant modulation of GW signals \cite{2017arXiv170204724D,2018PhRvD..98j4029D}.

%2.作为引力波天文学的一种重要工具，透镜化GWs的观测具有极大的研究价值。某些天体（例如暗物质晕[21-23]、中等质量黑洞[18-20]、 宇宙弦[24-26]）既不发光也不发射中微子，无法通过常规手段探测到。但是如果它们做为透镜体对GW产生影响，那么借助对这种透镜化GWs的探测，可能会为此类天体的观测和研究提供一种有效的方法，甚至可能发现一些新的天体。尤为重要的是，透镜化GWs的探测还有望为致密暗物质天体的存在提供直接证据[21, 22, 27-36]。不仅如此，这种效应还有望为我们提供一些更为精确的宇宙学测试。
As an important tool in GW astronomy, the observation of lensed GW signals holds great research value. 
Some celestial bodies (such as dark matter halos \cite{Choi_2021,Guo_2022}, intermediate-mass black holes \cite{Lai_2018,Gais_2022,Meena_2024} and cosmic strings \cite{Fern_ndez_N_ez_2016,Fern_ndez_N_ez_2017}) neither emit light nor neutrinos, thus cannot be detected by conventional observational means. 
However, if GWs are lensed by them, the detection of such lensed signals may offer an effective method for the observation and study of these celestial bodies, and even lead to the discovery of new ones.
Particularly important is that the detection of lensed GW signals is expected to provide a direct probe for the nature of dark matter \cite{Liu_2019,Jung_2019,Urrutia_2021,Wang_2021,Guo_2022,2022A&A...659L...5C,Fairbairn_2023,2023PhRvD.108l3543C,2023PhRvD.108j3529T,2024PhRvD.109l4020C,jana2024probingnaturedarkmatter}.
Not only those, this effect is also expected to provide us with some more precise tests in cosmology \cite{2019NatSR...911608C,2022ApJ...927...28L,2023PhRvD.108b4052G,2023JCAP...08..003H,2025RSPTA.38340134S}.

%3.近来，多项研究已经开展了对GW透镜效应的多次搜索[ ]，但均未明确搜索到透镜化GWs。其中使用的方法包括传统的基于匹配滤波的搜寻法和引入深度学习技术的搜寻法。尽管如此，随着越来越多的GW事件预计将被报道,加上关于探测率的预测[ ]仍让我们期待在未来观测运行中能够探测到受引力透镜效应影响GWs。
Recently, multiple studies have conducted numerous searches for the GW lensing effect, but none have clearly detected lensed GW signals. 
The methods employed include traditional search techniques based on matched filtering \cite{Abbott_2021,2023searchlensingsignaturesobserving,Janquart_2023,2025ApJ...980..258B,2024PhRvD.109b3028G} and those incorporating deep learning technology \cite{2021PhRvD.104l4057G,Kim_2022,2023PhRvD.107b3021S}. 
Nevertheless, as more and more GW events are expected to be reported, and given the predictions of detection rate \cite{2018MNRAS.476.2220L,PhysRevD.97.023012,2021PhRvD.103j4055W,2022ApJ...929....9X,2025PhRvD.111b4068B}, we remain hopeful that GW signals affected by the gravitational lensing effect will be detectable in future observational runs.

%4.由于波长的区别，GW的引力透镜效应与电磁波（EMW）的引力透镜效应有所不同。EMW的波长通常远小于透镜体的施瓦西半径，一般可以直接使用几何光学近似。目前地面GW探测器的观测频率范围较低，对于LVK网络探测到的GWs，其波长通常比质量M_{\rm L}<10^3 M_⨀的透镜体对应的施瓦西半径大，导致不可忽略的波效应。所以，在研究透镜化GW时需要充分考虑几何光学近似是否有效。
Due to the difference in wavelength, the gravitational lensing effect of GWs is distinct from that of electromagnetic waves (EMWs). 
The wavelength of EMWs is usually much smaller than the Schwarzschild radius of the lensing object, and the geometric optics (GO) approximation can generally be directly applied. 
The observation frequency range of ground-based GW detectors is very low, compared with the typical frequencies of EMWs. 
For GW signals detected by the LVK network, their wavelengths are usually larger than the Schwarzschild radius corresponding to lensing objects with mass \(M_{\rm L} \lesssim 10^{3}\) ${\rm M}_\odot$, leading to non-negligible wave effects \cite{2020MNRAS.492.1127M,Ali_2023}.
Therefore, when studying lensed GWs, it is necessary to fully consider whether the GO approximation is valid \cite{2004A&A...423..787T}.

%5.GW受到不同质量的透镜体影响时会产生不同的的波形变化。当透镜体的质量极大时，会出现间隔数分钟甚至数月的重复GW事件，从而形成多个GW源的“像”，这是强引力透镜效应。与之相比，由恒星或致密天体引起的微引力透镜效应，其重复事件间隔仅数毫秒到数秒。在微引力透镜效应下，GW传播中由于受透镜引力势影响产生的时间延迟和波的周期相当，不仅使得信号放大，还能在波形中产生“拍频”模式。利用微引力透镜化GWs作为探针，我们将可以对作为透镜体的恒星、小质量黑洞等致密天体开展更全面的研究。
When GWs are affected by lenses of different masses, they will undergo different waveform changes. 
When the mass of the lens is extremely large, repetitive GW events will occur at intervals of several minutes or even months, forming multiple ``images" of GW sources, which is the strong gravitational lensing effect \cite{2010PhRvL.105y1101S}. 
In contrast, the millilensing \cite{Liu_2023,Janquart_2023} effect caused by stars or compact objects has repetitive event intervals of only a few milliseconds to a few seconds. 
Under the millilensing effect, the time delay caused by the gravitational potential of the lens during the propagation of GWs is comparable to the period of the wave, which not only amplifies the signal but also generates a
``beating" pattern \cite{2022JCAP...07..022B} in the waveform.
By using millilensed GW signals as probes, we will be able to conduct more comprehensive studies on compact objects such as stars and small-mass black holes that act as lenses.

%6.对透镜化GW事件的参数推断是研究GW透镜效应的重要一步。传统的基于贝叶斯方法的参数推断流程极为耗时，特别对于增加了透镜体参数的透镜化GW事件，其效率更是极大降低。最近的研究表明，条件变分自编码器（CVAE）[75]、归一化流[76]等深度学习技术可以很好地快速实现参数推断，这为实现GWs的低延迟搜索提供了强大的工具。在文章[2]中研究了利用CVAE对微透镜化引力波事件进行参数推断，实现了在y（表征源的位置参数）和时间延迟∆t_d两参数上的快速参数推断，最快推断速度较基于bilby的方法快了五个数量级。本研究着眼于微引力透镜事件，利用点质量透镜模型来描述透镜对GW的影响，研究利用NSF进行13维参数（同时包含源、透镜体的信息：透镜体的质量M_{\rm L}、角直径距离D_{\rm L}、红移值Z_{\rm L}，源的啁啾质量M_C、质量比q、光度距离D_{\rm L}、红移值Z_{\rm S}、聚并相位ϕ_C、聚并时间t_{\rm c}、轨道倾角ι、赤经ra、赤纬dec）上的参数推断。
Parameter inference for lensed GW events is a crucial step in studying the GW lensing effect. 
The traditional parameter inference process based on Bayesian methods is extremely time-consuming, especially for lensed GW events with additional lens parameters, which significantly reduces its efficiency. 
Recent studies have shown that deep learning techniques such as Conditional Variational Autoencoders (CVAE) \cite{Gabbard_2021CVAE} and Normalizing Flows (NFs) \cite{Green:2020dnx,Green_2020NSF,Dax:2021tsq,2022MLS&T...3a5007S,2022PhRvD.105l4021W,2023PhRvL.130q1403D,2023CmPhy...6..212Z,2024arXiv240714298S,2024SCPMA..6730412D,2024ChPhC..48d5108S,2024PhRvD.109f4056L,2025Natur.639...49D,Xiong_2025} can effectively and rapidly perform parameter inference, providing powerful tools for low-latency GW signal searches. 
In Ref. \cite{Nerin_2025cvae}, the use of CVAE for parameter inference of
lensed GW events was investigated, achieving rapid parameter inference in two parameters, source's dimensionless location parameter \(y\) and time delay \(\Delta t_{\rm d}\), with the fastest inference speed being $5$ orders of magnitude faster than the Bayesian methods such as {\tt Bilby}.
In this work, we focus on millilensing events, using a point-mass lens model to describe the lensing effect on GWs, and explore the use of neural spline flows (NSFs) for parameter inference in 11 dimensions (including both sources and lenses information, i.e., lens mass \(M_{\rm L}\), angular diameter distance of lens \(D_{\rm L}\), source chirp mass \(\mathcal{M}_{\rm c}\), mass ratio \(q\), luminosity distance of source \(d_{\rm L}\), location parameter of the source \(y\), merger phase \(\phi_{\rm c}\), merger time \(t_{\rm c}\), orbital inclination \(\iota\), right ascension \(ra\), and declination \(dec\)).
%lens mass \(M_{\rm L}\), angular diameter distance \(D_{\rm L}\), redshift \(Z_{\rm L}\), source chirp mass \(\mathcal{M}_{\rm c}\), mass ratio \(q\), luminosity distance \(D_{\rm L}\), redshift \(Z_{\rm S}\), merger phase \(\phi_{\rm c}\), merger time \(t_{\rm c}\), orbital inclination \(\iota\), right ascension \(ra\), and declination \(dec\). 

%引力波源（例如双黑洞合并）通常具有自旋效应[]，研究其自旋有重要意义[]。然而，传统贝叶斯方法在处理含自旋参数的波形时面临显著挑战[]。利用NSF网络的快速性[]和泛化能力[]也可为解决这一困难提供一种方法。
GW sources (such as the merger of binary black holes) typically exhibit spin effects \cite{2011PhRvL.106x1101A,2022ApJ...937L..13C,2022ApJ...932L..19B,2022ApJ...935L..26F}, and studying their spins is of great significance \cite{2021PhRvD.104h3010R,2021PhRvL.126q1103B,2022PhRvD.106j3019T}.
However, traditional Bayesian methods encounter significant challenges when dealing with waveforms containing spin parameters \cite{2015PhRvD..91d2003V,2016PhRvD..94b4012H,2025PhRvD.111d4016S}. 
Utilizing the rapidity \cite{2024SCPMA..6730412D,2024ChPhC..48l5107W} and generalization ability \cite{2021PhRvD.104f4046C,2021PhRvD.103b4040X,2024MLS&T...5a5046W,2024PhLB..85839016X,Wang:2024oei} of the NSF network can also provide a solution to this difficulty.

%7.为此，我们在一组模拟的微引力透镜化GWs数据集上训练了一个NSF模型，使其能够学习到基于GW时序信号的透镜系统参数的概率潜在表示。结果表明，该模型可以较好实现对微引力透镜化GW事件的后验推断。同时，我们还证明了该模型对于训练时未引入的自旋参数也有较好的泛化性，这进一步拓展了该模型的适用范围。
In this work, we trained an NSF network on a set of simulated millilensed GW datasets, enabling it to learn the probabilistic latent representation of the lensing system parameters based on the GW time series signals. 
The results show that the network can effectively perform posterior inference for millilensed GW events.
Additionally, we demonstrate that the network has good generalization for spin parameters not introduced during training, further expanding the applicability of the network.

%8.本文的组织如下：第II节详细介绍我们工作所用到的方法，包括引力透镜理论的回顾、网络的配置以及数据的生成部分。第III节展示取得的的结果。最后，第IV节总结本工作的结论，并提出了改进方向和下一步的研究计划。
The organization of this paper is as follows.
Section ~\ref{sec2} provides a detailed introduction to the methodology used in our work, including a review of the gravitational lensing theory, the configuration of the network, and the data generation part.
Section ~\ref{sec3} presents the results. 
Finally, Section ~\ref{sec4} summarizes the conclusions of this work and proposes improvement directions and the next research plan.

\section{METHODOLOGY}\label{sec2}
\subsection{The theory of gravitational lensing}\label{sec2.1}

\begin{figure*}[!htp]
\includegraphics[width=0.8\textwidth, height=6cm]{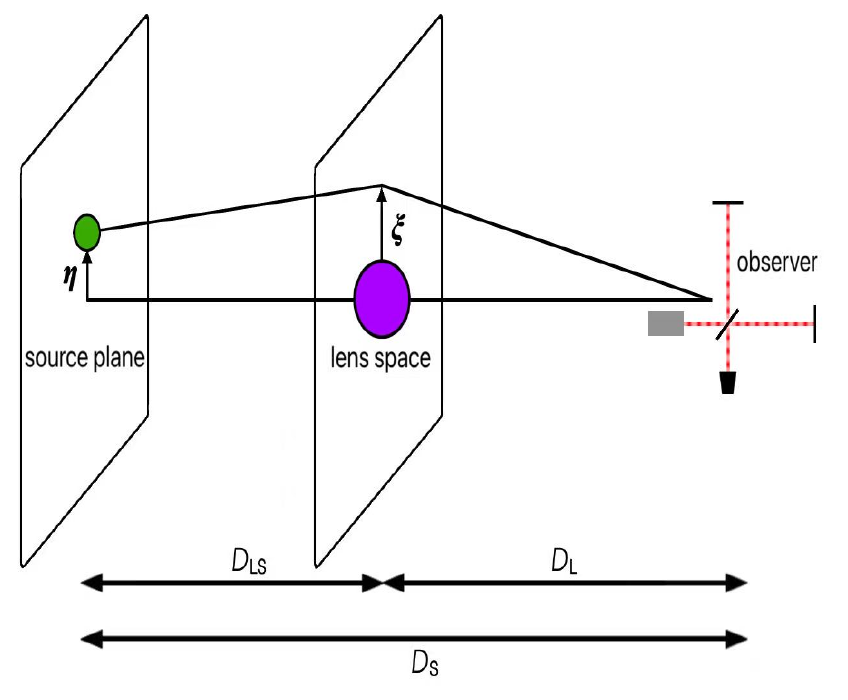}
\caption{\label{fig1} A classical gravitational lensing system composed of a GWs source on the source plane, a lens on the lens plane, and the observer. \(D_{\rm L}\), \(D_{\rm LS}\), \(D_{\rm S}\) are angular diameter distances from observer to lens, from lens to source, and from source to observer, respectively. The vector \(\bm{\eta}\) represents the GWs source's location. The vector \(\bm{\xi}\) is the impact parameter of the lens on GWs.}
\end{figure*}

%1.在引力透镜效应中，由于GW受到被称为透镜体的天体的影响，我们将观察到GWs的调制。如图1所示，当源位于由透镜和观察者的中心确定的轴线附近时，就会发生这种效应。由于透镜和源的尺寸范围远小于源到观察者和透镜到源的距离，所以可以将它们分别投影到透镜平面和源平面上[84]进行分析。D_{\rm L},D_{\rm L}S,D_S分别为观察者到透镜、透镜到源、源到观察者的角直径距离。向量η为GW源在源平面上相对于轴线的位置参数，向量ξ为透镜体对传播到透镜平面的GW的影响参数。在这种薄透镜近似下，GW仅在透镜平面上被由透镜体本身性质决定的二维引力势ψ(x)影响。这种影响最终体现在依赖于透镜模型的参数传输因子F(f)上。其中，F(f)=h ̃^L (f)/h ̃(f)将透镜化的波形h ̃^L (f)和未透镜化的波形h ̃(f)联系起来。所以，在波形数据模拟阶段，我们只需要先分别计算h ̃(f)和F(f)，再利用二者相乘便可生成透镜化GW的波形数据。
In the gravitational lensing effect, due to the impact of a celestial body known as lens on GWs, we will observe the modulation of signals.
As shown in Fig.~\ref{fig1}, this effect occurs when the GW source is located near the axis determined by the centers of lens and observer. 
Since the size range of lens and source is much smaller than the distances from source to observer and from lens to source, they can be projected onto the lens plane and the source plane respectively \cite{schneider1992gravitational} for analysis.
\(D_{\rm L}\), \(D_{\rm LS}\), and \(D_{\rm S}\) are angular diameter distances from observer to lens, from lens to source, and from source to observer, respectively. 
The vector \(\bm{\eta}\) represents the location of the source relative to the axis on the source plane, and the vector \(\bm{\xi}\) represents the impact parameter of lens on the GW propagating to the lens plane. 
Under this thin-lens approximation, GWs are only affected by two-dimensional gravitational potential \(\psi(x)\) determined by the properties of lens on the lens plane. 
This impact is ultimately reflected in the transmission factor \(F(f)\) dependent on the lens model.
Here, \(F(f) = \tilde{h}^L(f)/\tilde{h}(f)\) connects the lensed waveform \(\tilde{h}^L(f)\) and the unlensed waveform \(\tilde{h}(f)\).
Therefore, in the waveform data simulation stage, we only need to calculate \(\tilde{h}(f)\) and \(F(f)\) separately first, and then multiply the two to generate the waveform data of the lensed GWs.

%2.对于轴对称透镜，向量η和ξ可以简化为标量η和ξ。在源平面上取无量纲标量y≡η/(ξ_0 D_S/D_{\rm L}  )，它可作为源的位置参数。其中ξ_0是透镜平面上的一个依赖于透镜模型的特征长度，在点质量透镜(PML)模型中选为爱因斯坦半径：ξ_0=√((4GM_{\rm L})/((1+Z_{\rm L})c^2 )∙(D_{\rm L} D_{\rm L}S)/D_S ).Z_{\rm L}为透镜处对应的红移值。PML模型将透镜体近似为质点，适用于透镜的物理尺寸远小于爱因斯坦半径的情况，比如黑洞、致密暗物质团块以及类似的紧凑天体。
For axisymmetric lenses, the vectors \(\bm{\eta}\) and \(\bm{\xi}\) can be simplified to scalars \(\eta\) and \(\xi\). 
On the source plane, a dimensionless scalar \(y \equiv \eta / ( \xi _0 D_{\rm S} / D_{\rm L} )\) is taken as the location parameter of the source, where \( \xi _0\) is a characteristic length on the lens plane that depends on the lens model. In the point mass lens (PML) model, it is chosen as the Einstein radius: 
\begin{equation}
\xi_0 = \sqrt{\frac{4 G M_{\rm L}}{c^2} \cdot \frac{D_{\rm L} D_{\rm {LS}}}{D_{\rm S}}}.
\end{equation} 
The PML model approximates the lens as a point mass and is applicable when the physical size of the lens is much smaller than the Einstein radius, such as black holes, dense dark matter clumps, and similar compact celestial bodies.

%3.引力透镜效应会在观察者处造成有一定时间延迟的重复出现的GW事件。对于PML模型，在几何光学近似条件下，将在观察者处产生两个GW源的“图像”[41]。此时，F(f)在[16]中被给出：
%F(f)=|μ_+ |^(1/2)-i|μ_- |^(1/2) e^(2πif∆t_d )
%其中
%μ_±=1/2±(y^2+2)/(2y〖(y^2+4)〗^(1/2) )
%∆t_d=2M_{\rm L}{y〖(y^2+4)〗^(1/2)+2 ln⁡[(〖(y^2+4)〗^(1/2)+y)/(〖(y^2+4)〗^(1/2)-y)] }
%分别为两“图像”的放大系数和时间延迟。由此，引力透镜效应的影响可由源的位置参数y和透镜体的质量M_{\rm L}确定，并进一步由传输因子F(f)来体现。
The gravitational lensing effect will cause repeated GW events with certain time delay at the observer's location.
For the PML model, under the GO approximation condition, two ``images" of the GW sources will be produced at the observer's location \cite{Ali_2023}. 
At this point, \(F(f)\) is given in Ref. \cite{2004A&A...423..787T} as:
\begin{equation}
F(f) = |\mu_+|^{1/2} - i|\mu_-|^{1/2}e^{2\pi if\Delta t_{\rm d}},
\end{equation}
where
\begin{equation}
\mu_{\pm} = \frac{1}{2} \pm \frac{y^2 + 2}{2y(y^2 + 4)^{1/2}},
\end{equation}
\begin{equation}
\Delta t_{\rm d} = \frac{2GM_{\rm Lz}}{c^3} \left\{ y(y^2 + 4)^{1/2} + 2 \ln \left[ \frac{(y^2 + 4)^{1/2} + y}{(y^2 + 4)^{1/2} - y} \right] \right\},
\end{equation}
 are the magnification factors and time delay for the two ``images" respectively. $M_{\rm Lz} = M_{\rm L}(1+Z_{\rm L})$ is redshift mass of the lens.
 Thus, the impact of the gravitational lensing effect can be determined
by \(y\) and \(M_{\rm L}\), and further reflected by transmission factor \(F(f)\).

\subsection{Network architecture}\label{sec2.2}

%1.本工作提出一种基于深度残差待征提取-神经样条流的参数推断网络，用于从模拟的多探测器透镜化GWs中高效重建13维物理参数的后验分布。网络由特征编码器与概率解码器级联构成。
This work proposes a parameter inference network based on deep residual feature extraction and NSFs for efficiently
reconstructing the 11-dimensional posterior distribution of physical parameters from simulated multi-detector lensed GW signals. 
The network is composed of a feature encoder and a probabilistic decoder cascaded together.

%2.特征编码器采用改进的ResNet-50,它的结构与文章[2]中提到的完全类似。其输入为四通道的时域信号，分别对应LIGO(H1/L1),Virgo(V1), KAGRA(K1)四个探测器所探测到的GWs。时域特征经过ResNet-50网络最终被映射为多维特征向量，并作为输入传输给概率解码器进行后验推断。
The feature encoder adopts an improved ResNet-50, whose structure is exactly the same as that described in Ref. \cite{2016cvpr.confE...1H}.
Its input is the four-channel time-domain signal, corresponding to the GW signals detected by the four detectors: LIGO (H1/L1), Virgo (V1), and KAGRA (K1).
The time-domain features are ultimately mapped into a 512-dimensional feature vector through the ResNet-50 network, and then it is transmitted as input to the probabilistic decoder for posterior inference.

%3.概率解码器采用神经样条流(NSFs)架构，具体而言是有理二次神经样条流（RQ-NSFs），该部分具体架构详见[ ]。归一化流（NFs）是一种使用深度学习技术的概率建模方法，用于学习复杂的概率分布。其核心思想是构建参数到因变量的可逆变换f_ϕ:R^13×R^512→R^13，从而获得简单基础分布（例如正态分布）与复杂后验分布之间的映射关系。它是一种能够通过网络变换准确计算概率密度函数的雅可比行列式的可逆神经网络，每个变换都根据输入数据和神经网络前一层的输出进行转换，从而预测并逐渐逼近后验分布。RQ-NSF通过在NF中引入单调有理二次样条，可以增强自回归变换的灵活性。它利用单调有理二次变换来替代回归层中的加法或仿射变换，在自回归变换之间，对样本分量进行反转，以确保变换序列能够专门且完整地转换所有分量。通过复合变换f_ϕ=f_N∘⋅⋅⋅∘f_1，NSF网络导出的参数后验分布可解析表示为：
%p(θ|x)=π(〖f_ϕ (θ;h(x))〗^(-1) )|det〖J_(f_ϕ )〗^(-1) |.其中x表示输入网络的信号数据，θ对应所要推断的参数。u=f(x)与x通过变换f联系在一起，pai(u)通常被选为标准正态分布。
%模型训练采用最大似然准则，优化目标为最小化负对数似然损失：
%L=-1/N ∑_(i=1)^Nlog⁡p(θ_i |x_i ) 
The probability decoder adopts the NSF architecture, specifically the Rational Quadratic Neural Spline Flows (RQ-NSFs) \cite{2019arXiv190604032D}. 
The NF is a probabilistic modeling method that utilizes deep learning techniques to learn complex probability distributions.
Its core idea is to construct a bijective transformation $f_x$ from parameters to dependent variables, thereby obtaining the mapping relationship between a simple base distribution (such as a normal distribution) and a complex posterior distribution. 
It is a reversible neural network capable of accurately calculating the Jacobian determinant of the probability density function through network transformations.
Each transformation is based on the input data and the output of the previous layer of the neural network, thereby predicting and gradually approximating the posterior distribution. 
RQ-NSF enhances the flexibility of autoregressive transformations by introducing monotonic rational quadratic splines in NFs. 
It uses monotonic rational quadratic transformations to replace the addition or affine transformations in the regression layer and reverses the sample components between autoregressive transformations to ensure that the transformation sequence can specifically and completely transform all components. 
Through the invertible and differentiable transformation \(f_x\), the parameter posterior distribution derived by the NSF network can be analytically expressed as \cite{rezende2015variational,DBLP:journals/corr/PapamakariosPM17}:
\begin{equation}
p(\theta|x) = \pi\left(f_x^{-1}(\theta)\right) \left| \det J_{f_x}^{-1} \right|,
\end{equation}
where $x$ represents the data input to the network, and $\theta$ corresponds to the parameters to be inferred. 
The $u = f_x^{-1}(\theta)$ is linked to $\theta$ through the transformation $f_x$, and $\pi(u)$ is usually chosen to be a standard normal distribution.

Network training is conducted using the maximum likelihood criterion, with the optimization objective being to minimize the negative log-likelihood loss:
\begin{equation}
\mathcal{L} = -\frac{1}{N}\sum_{i=1}^{N}\log p(\theta_i|x_i).
\end{equation}

\begin{figure}[!htp]
\centering
\includegraphics[width=0.4\textwidth, height=8cm]{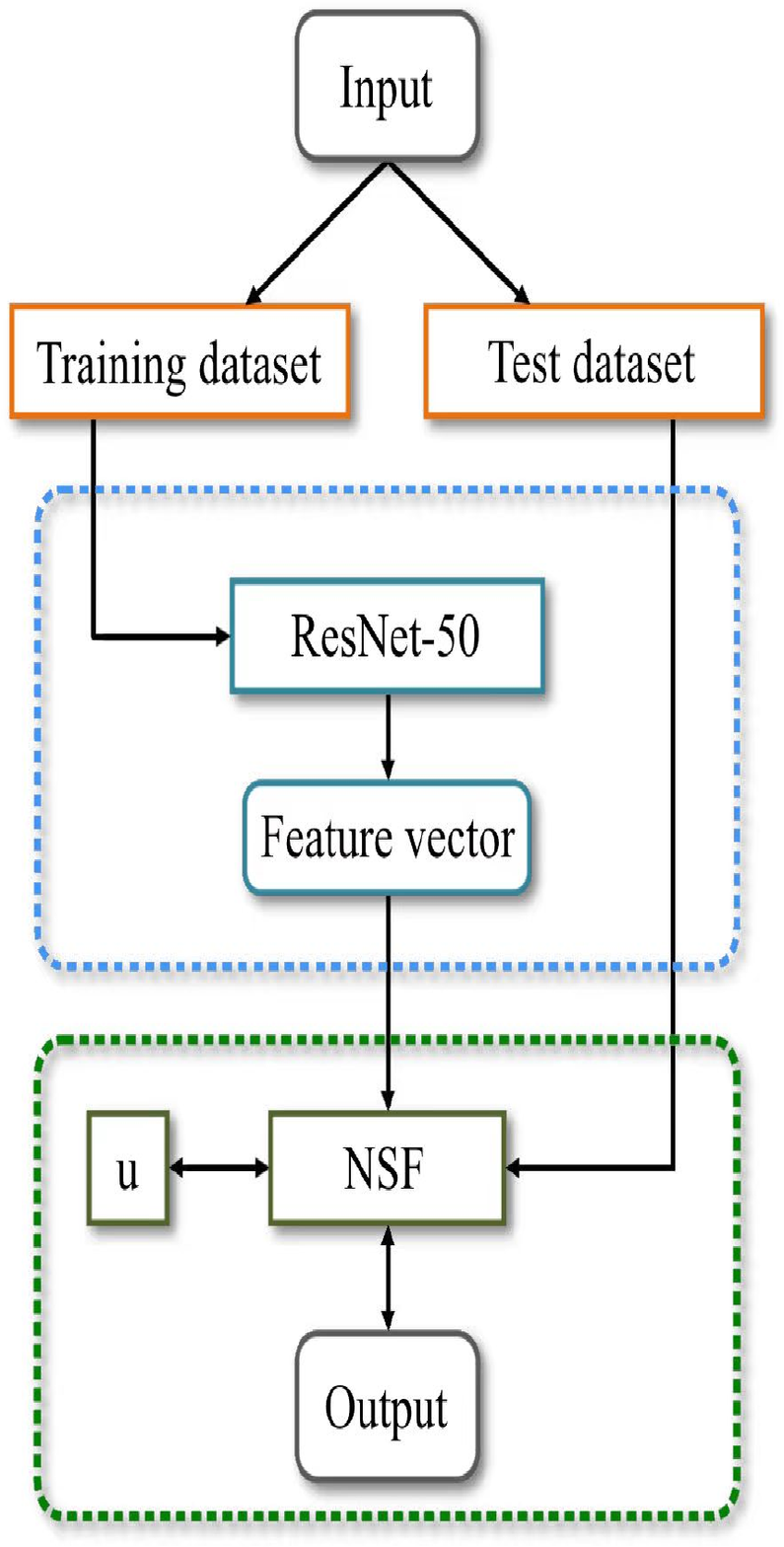}
\centering \caption{\label{fig2}Workflow of the NSF network. The input is the GW time series signal after lensing. The signal passes through the ResNet-50 network (that is the part enclosed by the blue dotted line) to extract feature vectors. Finally, the extracted feature vectors are inputted into the NSF network (that is the part enclosed by the green dotted line) to obtain posterior inference of parameters.}
\end{figure}

%4.在训练过程中，我们使用AdamW优化器，初始学习率为0.0002，学习率衰减因子为 0.00002，批量大小为20，隐藏层大小为 1024。为确保归一化流网络预测的有效性并防止过拟合，并采用 50 轮早停策略。我们还在每个训练周期开始时重新生成10,000个新数据点，其中80%用于训练，20%用于验证。络的训练流程如图2所示。
The NSF network's key parameters are shown in Table \ref{tab1}.
During the training process, to ensure the validity of the predictions made by the NSF network and prevent overfitting, we employ the AdamW optimizer and adopt a 50-round early stopping strategy.
We also regenerate $1.2 \times 10^{5}$ new data at the beginning of each training cycle, with $10^5$ used for training and $2 \times 10^4$ for validation. 
The training process of the network is illustrated in Fig.~\ref{fig2}.
Our entire network is trained on an NVIDIA RTX A6000 GPU with 48 GB of memory.
%see Table \ref{tab1}

\begin{table}
\caption{The key parameters of the NSF network.}\label{tab1}
\centering
\setlength\tabcolsep{15pt}
\renewcommand{\arraystretch}{1.5}
\begin{tabular}{cc}
        \hline \hline
        Parameter & Value \\
        \hline 
        Activation function & ReLU \\
        Number of bins & 8 \\
        Flow steps & 3 \\
        Transform blocks & 5 \\
        Batch size & 20 \\
        Learning rate decay factor & $2\times10^{-5}$ \\
        Initial learning rate & $2\times10^{-4}$ \\
        Hidden layer size & 1024 \\
        \hline \hline
\end{tabular}
\end{table}

\subsection{Data sets}\label{sec2.3}
%\section{Data set}\label{sec4}

%1.模型的训练集和验证集都由参数集和对应的GW时域信号组成。数据基于平坦ΛCDM宇宙学模型（c=299792.458km/s，H_0=70 km/s/Mpc，Ω_m=0.3， Ω_Λ=0.7），选取适当的先验参数（见表1）生成。
The training, validation and test sets of the network are composed of parameter sets and the corresponding GWs time-domain signals.
The data are based on a flat $\Lambda$CDM cosmological model ($H_0 = 70$ km/s/Mpc, $\Omega_m = 0.3$, $\Omega_\Lambda = 0.7$), and are generated using appropriate prior parameters (see Table \ref{tab2}).        

%用来生成GW波形数据的各参数的先验范围。为了研究方便，表中未提到的参数我们将其简单设置为零。只有在研究网络对自旋参数的泛化性时，我们改变了自旋值以生成一系列含自旋的测试集。
\begin{table*}[t]
\caption{The prior ranges of each parameter used to generate GWs waveform data. For the convenience of the research, the parameters not mentioned in the table have been simply set to zero.
Only when studying the generalization ability of the network for spin parameters, we change the spin values to generate a series of test sets containing spins.}
\label{tab2}
\setlength{\tabcolsep}{10pt} % Slightly relaxed column spacing
\renewcommand{\arraystretch}{1.5} % Improve row height for readability
\centering
\begin{tabular}{ccccc}
\hline\hline
\textbf{Parameter} & \textbf{Description} & \textbf{Range} & \textbf{Unit} & \textbf{Distribution} \\
\hline
$\mathcal{M}_{\rm c}$ & Chirp mass of the binary system & $(5, 80)$ & $\mathrm{M}_\odot$ &  LogUniform\\
$q$ & Ratio of masses of the two black holes & $(0.125, 1)$ & $\cdots$ & Uniform \\
$M_{\rm L}$ & Mass of the lens  & $(10^3, 10^5)$ & $\mathrm{M}_\odot$ & LogUniform \\
$Z_{\rm L}$ & Redshift value at the lens & $(0, 1)$ & $\cdots$ & PowerLaw \\
$Z_{\rm S}$ & Redshift value at the source & $(Z_{\rm L}, 1)$ & $\cdots$ & PowerLaw \\
$\delta$ & Ecliptic latitude of the binary system & $\left(-\frac{\pi}{2}, \frac{\pi}{2}\right)$ & $\mathrm{rad}$ & Cosine \\
$\alpha$ & Ecliptic longitude of the binary system & $(0, 2\pi)$ & $\mathrm{rad}$ & Uniform \\
$\psi_{\rm s}$ & Polarization angle of the gravitational wave & $(0, \pi)$ & $\mathrm{rad}$ & Uniform \\
$\iota$ & Angle of inclination of the binary orbit & $(0, \pi)$ & $\mathrm{rad}$ & Sine \\
$\phi_{\rm c}$ & Phase at the moment of coalescence & $(0, 2\pi)$ & $\mathrm{rad}$ & Uniform \\
$t_{\rm c}$ & Time of coalescence & $(9.0, 9.1)$ & $\mathrm{s}$ & Uniform \\
$\eta$ & Location of the binary system & $(10^{-6}, 0.5)$ & $\mathrm{pc}$ & Uniform \\
\hline\hline
\end{tabular}
\end{table*}

%2.另外，源到观察者的光度距离 D_{\rm L}和透镜到观察者的角直径距离D_{\rm L}的先验可由Z_{\rm L}得到，且都满足幂律分布。Z_{\rm L}和Z_{\rm S}的概率密度表达式分别为 p(Z_{\rm L} )=3Z_{\rm L}^2、p (Z_{\rm S}│Z_{\rm L} )=3/(1-Z_{\rm L}^3 ) Z_{\rm S}^2。其中，为了更好覆盖参数空间，实现在不同数量级上均匀采样，选取的M_{\rm L}和M_C符合对数均匀分布。M_{\rm L}、y范围的选取主要参考文献[]。在[]中表明，当透镜体的红移质量M_{\rm L}Z与GWs的频率f使w=8πM_{\rm L}z f~1 时，传输因子F(f)会逐渐收敛到几何光学近似极限。对于我们考虑的M_{\rm L}>10^3和f>20Hz，对应的w>2.5，几何光学近似在一定范围内是可取的。虽然直接采用F的解析形式[]才是最为严谨的做法，但出于验证性探索的出发点，以及计算成本与实用性的考虑，我们选取了更为简洁的几何光学近似结果。关于引力波的透镜效应中几何光学近似的适用范围的相关讨论可见[]。
In addition, the prior of \(d_{\rm L}\) from the source to observer and \(D_{\rm L}\) can be obtained from \(Z_{\rm L}\), and both follow a power-law distribution. 
The probability density function for $Z_{\rm L}$ and $Z_{\rm S}$ are as follows: $p(Z_{\rm L}) \propto 3Z_{\rm L}^2$, and $p(Z_{\rm S}|Z_{\rm L}) \propto 3/(1 - Z_{\rm L}^3)Z_{\rm S}^2$.
To better cover the parameter space and achieve uniform sampling at different
orders of magnitude, the selected \(M_{\rm L}\) and \(\mathcal{M}_{\rm c}\) follow a log-uniform distribution. 
The selection of $M_{\rm L}$ and $\eta$ range mainly refers to Ref. \cite{Kim_2022}.
It is indicated in Ref. \cite{2004A&A...423..787T} that when $M_{\rm Lz}$ and the frequency $f$ of GW satisfy the condition that \(\omega \equiv 8 \pi GM_{\rm Lz} f/c^3 \gtrsim 1\), transmission factor \(F(f)\) in the wave optics (WO) asymptotically converges to the GO limit.
For the cases $M_{\rm L} > 10^3$ ${\rm M}_\odot$ and $f > 20$ ${\rm Hz}$ considered in our study (correspondingly $\omega > 2.5$), GO approximation is applicable within a certain range.
Although the most rigorous approach would be to directly adopt the analytical form of $F(f)$ \cite{2004A&A...423..787T}, for the purpose of exploratory verification and considering the computational cost and practicality, we chose a more concise GO approximation result. 
For discussions on the applicable range of the geometric optical approximation in the lensing effect of GWs, refer to Refs. \cite{2020MNRAS.492.1127M,2022JCAP...07..022B,Bondarescu_2023,2003ApJ...595.1039T}.

%3.双黑洞（BBH）并和GW波形数据借助Pycbc软件包生成，并利用第II节中所述理论模拟波形的透镜化调制。首先模拟生成透镜化的GW时序信号。由上述物理参数的采样范围，利用Pycbc中的get_td_waveform方法生成IMRPhenomTPHM近似模型下的时域波形h_+ (t),h_× (t)，使用antenna_pattern方法生成LIGO（H1/L1）、Virgo（V1）、KAGRA（K1）四个探测器各自的极化系数F_+,F_×，再由h ̃(f)=F_+∙h_+ (t)+F_×∙h_× (t)得到各探测器对应的非透镜化GW波形h ̃_H1 (f),h ̃_L1 (f),h ̃_V1 (f),h ̃_K1 (f)。按照第II节中所述方法计算传输因子F(f)，即可生成相应的透镜化GW波形h ̃_H1^L (f),h ̃_L1^L (f),h ̃_V1^L (f),h ̃_K1^L (f)。其中加入了限制条件时间延迟∆t_d≤0.2s和通量比I=|μ_- |/|μ_+ | ∈[0,1]，前者用于避免因延迟过长导致信号超出观测窗口，后者约束次像与主像的强度比，用于筛选符合物理情景的透镜事件。然后为GWs注入噪声。利用noise_from_psd方法，基于各探测器O4观测期的噪声灵敏度曲线生成了四段长度为10s，生成频率范围为20Hz~4096Hz的高斯噪声。即高通频率为20Hz，采样频率为4096Hz。其中探测器的噪声灵敏度曲线利用aLIGO140MpcT1800545方法计算。最后，将透镜化的GWs注入噪声信号中，限制信噪比SNR∈[10,50]，并白化处理，便得到了长度为10s的透镜化的GW波形数据。
The GW waveform data from the binary black hole (BBH) mergers are generated using the {\tt Pycbc} software package, and the lensing modulation of the theoretical simulated waveforms is carried out as described in Section ~\ref{sec2.1}. 
Initially, the lensed GW time-series signal is modeled.
Based on the sampling range of the above physical parameters, the unlensed GW waveforms \(\tilde{h}_{H1}(f)\), \(\tilde{h}_{L1}(f)\), \(\tilde{h}_{V1}(f)\), \(\tilde{h}_{K1}(f)\) under the \texttt{IMRPhenomTPHM} \cite{2022PhRvD.105h4040E} approximation model for each of the four detectors are generated using the {\tt Pycbc}.
Then, we calculate the transmission factor \(F(f)\) as described in Section ~\ref{sec2.1}, and the corresponding lensed GW waveforms \(\tilde{h}_{H1}^L(f)\), \(\tilde{h}_{L1}^L(f)\), \(\tilde{h}_{V1}^L(f)\), \(\tilde{h}_{K1}^L(f)\) can be generated.
During the process of data generation, the time delay \(\Delta t_d \leq 0.2 \ 
~\text{s}\) and the flux ratio \(I = {|\mu_-|}/{|\mu_+|} \in [0, 1]\) are added as constraints. 
The former is used to prevent the signal from exceeding the observation window due to excessive delay, while the latter constrains the intensity ratio of the secondary image to the primary image, which is used to screen lensing events that conform to the physical scenario.
Next, Gaussian noise is injected into the GW signals.
Leveraging the \texttt{noise\_from\_psd} method, four segments of Gaussian noise with a length of 2 s and a frequency range of \(20 \, \text{Hz} \sim 4096 \, \text{Hz}\) are generated based on the noise sensitivity curves of each detector during the O4 observing run.
That is, the high-pass frequency is \(20 \, \text{Hz}\) and the sampling frequency is \(4096 \, \text{Hz}\). 
Finally, the lensed GW signals are embedded within the noisy background with a signal-to-noise ratio SNR \(\in [10, 50]\), and then whitened, resulting in 2 s of lensed GW waveform data.

%4.训练集、验证集和测试集使用相同的方法生成,且这些数据集互不相交。正常的训练集和测试集均在无自旋的情况下生成，为了测试网络对于自旋参数的泛化性，我们还利用相同的方法生成了一系列具有不同自旋值的其它测试集。将参数集样本输入网络进行训练之前，我们使用最大-最小归一化方法将其归一化到了[-1,1]范围内。
The training set, validation set and test set are generated using the same method, and these datasets are mutually exclusive. 
Both the normal training set and the test set are generated without spin. 
To test the network's generalization ability for spin parameters, we also generate a series of other test sets with different spin values using the same
method. 
Before inputting the parameter set samples into the network for training, we normalize them to the range of \([-1, 1]\) using the max-min normalization method.

\section{Results and discussion}\label{sec3}

\begin{figure}[!htp]
\centering
\includegraphics[width=0.5\textwidth]{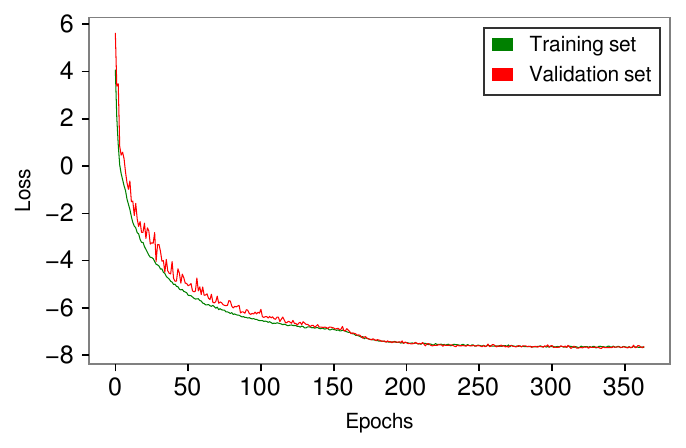}
\centering \caption{\label{fig3}Changes in loss values on the training set and validation set during training.
The NSF network is finally converged quite well.}
\end{figure}

%1.我们使用第III节中描述的架构训练NSF网络，在316个周期后实现了损失函数的平稳演变，如图3所示。损失值最终收敛至-11.8，训练损失和验证损失始终接近，没有过拟合的迹象。在训练结束时，我们保存性能最佳的那一个时期的模型，即验证损失最低的那个时期的模型，并将其用于测试。 
We train the NSF network using the architecture described in Section ~\ref{sec2.2}. 
After $319$ epochs, the loss function achieves a stable evolution, as shown in Fig.~\ref{fig3}. 
The loss value of training set eventually converges to about $-7.7$, and the training loss and validation loss remain close throughout, with no sign of overfitting.
At the end of training, we save the network of the epoch with the best performance, that is, the one with the lowest validation loss, and use it for testing.

\begin{figure}[!htp]
\centering
\includegraphics[width=0.5\textwidth]{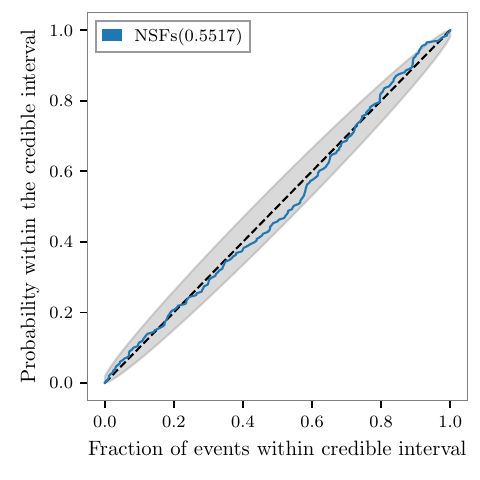}
\centering \caption{\label{fig4}Joint KS test based on the aggregated posterior quantiles of 11 parameters' marginalized distribution.
This curve is obtained using 200 samples, and the dotted black diagonal line represents the ideal result.
The legend indicates the corresponding $p$-value, and the gray area represents the $95\%$ confidence interval.
It demonstrates that the parameter inference accuracy of the NSF network is quite satisfactory.
}
\end{figure}

%2.我们随机从测试集中选取了200起事件，以验证NSF网路对于选定的11维参数的推断的稳健性。通过注入200个模拟的透镜化GW波形，我们测量了在一系列置信水平下真实参数的频率。我们还计算了每个参数后验分布的百分位数在参数注入值处的累积概率，然后在图4中绘制了参数联合分布的累积分布函数（CDF）。检验基于 200 个测试事件的联合后验标量全局秩进行，对每个事件计算单一全局秩后聚合所有结果并补充 0 和 1 边界值，通过 11 维联合后验概率天然考虑参数间的相关性，对全部参数执行联合检验，最终将全局秩的经验分布与均匀分布做双样本 KS 检验。理想情况下，注入的参数 CDF 应该呈现出对角线模式，因为CDF表示各个事件分布的累积和。曲线与对角线的接近程度表明参数推断的准确性。如图，曲线与对角线紧密贴合，整体位于95%的置信区间内，这表明NSF网路在准确估计参数方面具有出色的表现，图例中还标注了该模型的后验分布中获得的p值。
We randomly select $200$ events from the test set to verify the robustness of the NSF network's inference for selected 11-dimensions parameters. 
By injecting $200$ simulated lensed GW waveforms, we measure the frequency of the true parameter over a range of confidence levels.
We calculate the cumulative probability of each parameter posterior’s percentile at the injection values of parameters, then plot the cumulative distribution function (CDF) for the joint distribution of the parameters in Fig.~\ref{fig4}.
The test is conducted based on the joint posterior scalar global rank of $200$ test events. 
For each event, a single global rank is calculated, and all results are aggregated and supplemented with $0$ and $1$ boundary values.
Through the 11-dimensional joint posterior probability, the correlation between parameters is naturally considered. 
The joint test is performed on all parameters, and finally, a two-sample KS test is conducted between the empirical distribution of the global rank and the uniform distribution.
Ideally, the injected parameter CDF should show a diagonal pattern because the CDF represents the cumulative sum of the distribution of each event. 
The closeness of the curve to the diagonal indicates the accuracy of the parameter inference. 
As shown in the figure, the curve closely follows the diagonal, and the entire curve is within the $95\%$ confidence interval, indicating that the NSF network perform excellently in accurately estimating parameters. 
The $p$-value obtained from the posterior distribution of the network is also marked in the legend.

%NSF网络最终较好地收敛了。

%图例中标注了多次测试后p值的平均值，灰色区域为95%的置信区间。它证明NSF网络的参数推断精度比较理想。
%11个参数边缘后验分位数聚合后的联合 KS 检验

%它展现了NSF对主要参数恢复的精度，以及对部分参数简并性的识别。
\begin{figure*}[!htp]
\centering
\includegraphics[width=1\textwidth]{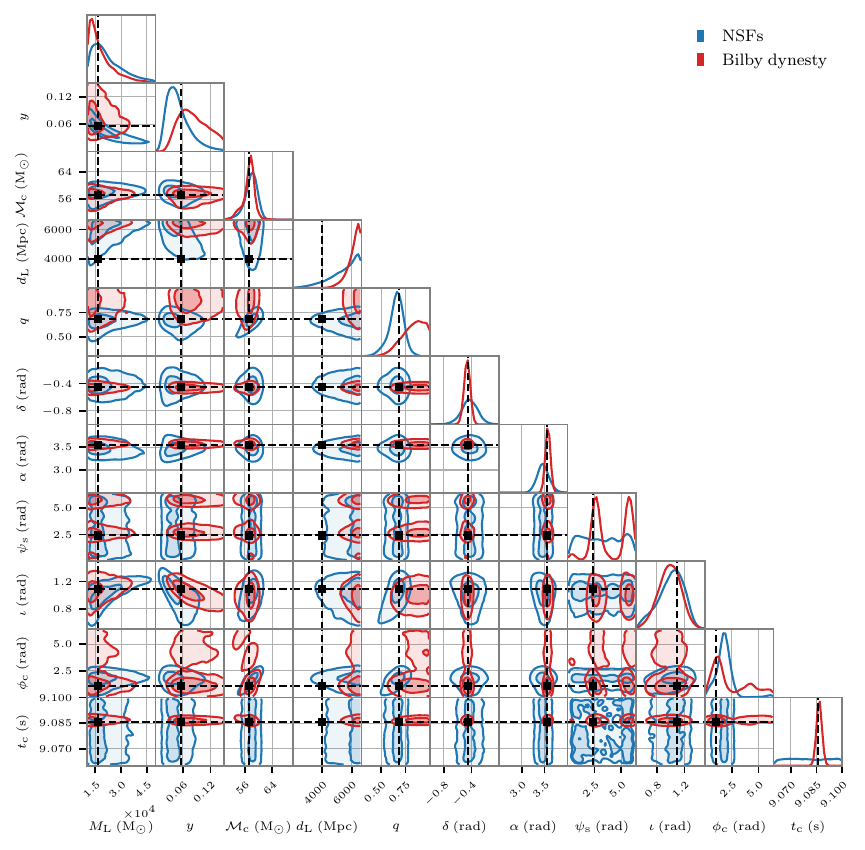}
\caption{\label{fig5}Marginal one-dimensional and two-dimensional posterior distribution plots for the same event without spin, inferred using the NSF network and {\tt Bilby dynesty}. 
The true values of the parameters are marked by black dashed lines. 
The blue curves represent the results from the NSF network, while the red curves represent the results from {\tt Bilby dynesty}.
In the two-dimensional distribution plots, the contour lines represent the $95.4\%$ and $68.3\%$ confidence regions, respectively.
It demonstrates the accuracy of NSFs in restoring the key parameters, as well as its ability to identify the degeneracy of some parameters.
}
\end{figure*}

%该表格量化展示了NSF和Bilby对Fig.5中的引力波事件的各个参数的推断结果。
\begin{table*}[t]
\caption{Comparison of the parameter inference results from NSFs and {\tt Bilby dynesty} for the gravitational-wave event shown in Fig.~\ref{fig5}. Reported values are given with their $1\sigma$ confidence intervals.}
\label{tab4}
\setlength{\tabcolsep}{10pt} % Slightly relaxed column spacing
\renewcommand{\arraystretch}{1.5} % Improve row height for readability
\centering
\begin{tabular}{cccc}
\hline\hline
\textbf{Parameter} & \textbf{Injected value} & \textbf{Bilby} & \textbf{NSFs} \\
\hline
$M_{\rm L}~({\rm M}_{\odot})$ & $1.66\times10^4$ & $8.70\times10^3{}^{+5.78\times10^3}_{-6.06\times10^3}$ & $2.18\times10^4{}^{+1.06\times10^4}_{-1.08\times10^4}$ \\
$y$ & $5.65\times10^{-2}$ & $0.16{}^{+0.05}_{-0.10}$ & $0.05{}^{+0.02}_{-0.03}$ \\
$\mathcal{M_{\rm c}}~({\rm M}_{\odot})$ & $57.20$ & $56.88{}^{+1.04}_{-1.53}$ & $57.43{}^{+1.72}_{-1.61}$ \\
$d_{\rm L}~(\rm Mpc)$ & $3.97\times10^3$ & $6.11\times10^3{}^{+3.93\times10^2}_{-4.13\times10^2}$ & $5.44\times10^3{}^{+9.49\times10^2}_{-1.02\times10^3}$ \\
$q$ & $0.68$ & $0.82{}^{+0.12}_{-0.12}$ & $0.64{}^{+0.07}_{-0.07}$ \\
$\delta~(\rm rad)$ & $-0.45$ & $-0.48{}^{+0.03}_{-0.03}$ & $-0.45{}^{+0.12}_{-0.11}$ \\
$\alpha~(\rm rad)$ & $3.55$ & $3.55{}^{+0.03}_{-0.03}$ & $3.52{}^{+0.07}_{-0.20}$ \\
$\psi_{\rm s}~(\rm rad)$ & $2.40$ & $3.88{}^{+1.89}_{-1.59}$ & $3.14{}^{+2.18}_{-2.14}$ \\
$\iota~(\rm rad)$ & $1.09$ & $0.97{}^{+0.16}_{-0.17}$ & $0.98{}^{+0.19}_{-0.20}$ \\
$\phi_{\rm c}~(\rm rad)$ & $1.05$ & $2.31{}^{+2.19}_{-1.61}$ & $1.72{}^{+0.49}_{-0.68}$ \\
$t_{\rm c}~(\rm s)$ & $9.09$ & $9.09{}^{+0.0008}_{-0.0008}$ & $9.05{}^{+0.03}_{-0.03}$ \\
\hline\hline
\end{tabular}
\end{table*}

%3.p 值为模型预测的累积分布与预期理论分布之间的一致性提供了定量度量，它基于柯尔莫哥洛夫-斯米尔诺夫（KS）检验[124]方法计算。在引力波参数估计中，KS 检验可用于评估估计的参数分布与预期理论分布之间的吻合程度 [127]。KS检验首先计算两个分布的CDF，然后根据 CDF 计算两个分布之间的距离，即KS统计量。基于KS统计量便可计算p值，通常被用定量评估两个分布之间的相似性或差异性。利用p值与显著性水平进行比较，如果 p 值大于显著性水平，说明被比较的两个分布在统计学上是相似的。这里，显著性水平取值为0.05。
The $p$-value provides a quantitative measure of the consistency between the cumulative distribution predicted by the network and the expected theoretical
distribution. 
It is calculated based on the Kolmogorov-Smirnov (KS) test \cite{lopes2011kolmogorov} method.
In GW parameter estimation, the KS test can be used to assess the degree of agreement between the estimated parameter distribution and the expected theoretical distribution \cite{Biwer_2019}. 
The KS test first calculates the CDF of the two distributions and then computes the distance which is called the KS statistic between them based on the CDF. 
The $p$-value can be calculated based on the KS statistic and is typically used to quantitatively evaluate the similarity or difference between two distributions. 
By comparing the $p$-value with the significance level, if the $p$-value is greater than the significance level, it indicates that the two compared distributions are statistically similar.
Here, the significance level is set at 0.05.
%{\color{red}$\mathcal{H}_0$: ??? $\mathcal{H}_1$: ??? }%学生收到，It has been resolved in next paragraph.

%4.我们通过比较两种分布，即理想分布和由NSF网络生成的参数联合分布，来评估网络建立的后验是否合理。理想分布是均匀分布，因为百分位数是在0到1之间均匀分布。另一种分布是每个参数后验分布的百分位数在参数注入值处的累积概率。我们的零假设是：假设这两种分布是相同的分布，这意味着NSF网络给出的参数后验推断是合理的；而备择假设自然是两种分布不相同，意味着网络给出的后验不合理。将算得的p值与 0.05 的显著性水平进行比较。如果p值小于0. 05，则认为零假设不成立。我们训练的NSF网络最终给出的p值=0.5517>0.05，这有力地说明网络对11个参数的整体推断是合理的。
We evaluate the reasonableness of the posterior established by the network by comparing two distributions, namely the ideal distribution and the joint distribution of parameters generated by the NSF network.
The ideal distribution is a uniform distribution because the percentiles are uniformly distributed between $0$ and $1$. 
The other distribution is the cumulative probability of the percentiles of each parameter's posterior distribution at the parameter injection value. 
Our null hypothesis is: assuming that these two distributions are the same, this means that the posterior inference of the parameters given by the NSF network is reasonable; while the alternative hypothesis naturally states that the two distributions are different, meaning that the posterior given by the network is unreasonable. 
The calculated $p$-value is compared with the significance level of $0.05$. 
If the $p$-value is less than $0.05$, the null hypothesis is considered invalid. the NSF network we trained ultimately gives a $p$-value $= 0.5517 > 0.05$, which strongly demonstrates that the overall inference of the network on $11$ parameters is reasonable.

%为何bilby对于天空参数推断很准确，而NSF差？是因为NSF不区分各个探测器，导致无法提取方位特征？
%5.在整体参数推断合理的基础上，我们又具体比较了两种方式对随机挑选的一起事件的推断效果。图5展示了我们训练的NSF网络和bilby dynesty对无自旋测试集样本中的随机一例透镜化GW事件的后验估计的比较，其中bilby dynesty的先验设置与表I相同。对于我们研究的11个参数，NSF网络仅用了不到1秒便成功完成了后验推断，并且对大多参数给出的后验分布都与bilby给出的结果接近。整体上NSF给出的二维后验分布比较弥散，我们认为这是因为网络没有很好提取IMRPhenomTPHM模型引入的高阶模式特征[]。与之相比，bilby dynesty由于引入波形的高阶模式，给出的二维后验分布确实更为集中，但耗费的时间较NSF增加了4个数量级。在我们十次的测试中，bilby dynesty的后验推断速度最快为64,091秒，而NSF网络的推断速度平均约为0.8秒。这使得NSF应用于实时引力波信号搜索成为可能。
On the basis of the reasonable inference of overall parameters, we further specifically compared the inference effects of the two methods.
Figure~\ref{fig5} and Table~\ref{tab4} show a comparison of the posterior inference of a randomly selected lensed GW event in the spinless test set sample between our trained NSF network and {\tt Bilby dynesty}, where the prior settings of {\tt Bilby dynesty} are the same as those in Table ~\ref{tab1}. For the 11 parameters we studied, the NSF network successfully completed the posterior inference in less than $1$ s and provided posterior distributions for most parameters that were close to those given by {\tt Bilby dynesty}. Here, $M_{\rm L}$ refers to the lens mass in the source frame (not the red-shifted mass). Overall, the two-dimensional posterior distribution given by NSFs is relatively diffuse. We believe this is because the network did not extract the high-order mode features \cite{2022PhRvD.105h4040E,2022ApJ...924...79E,2022MNRAS.517.2403M,2025arXiv250409679D} introduced by the \texttt{IMRPhenomTPHM} model well. In contrast, due to the introduction of higher-order modes of the waveform, the two-dimensional posterior distribution given by {\tt Bilby dynesty} is indeed more concentrated, but the time it took is four orders of magnitude longer than that of NSFs. In our $10$ tests (see Table \ref{tab3}), the posterior inference speed of {\tt Bilby dynesty} is the fastest at $60070$ s, while the inference speed of the NSF network averaged approximately $0.8$ s. This makes it possible for NSFs to apply in real-time GW signals search.

%在十次测试下，Bilby dynesty和NSF网络分别独立完成13维参数后验推断的时间统计结果。
\begin{table}
\caption{The statistical results of the time taken for {\tt Bilby dynesty} and NSFs network to independently complete the posterior inference of 11-dimensional parameters, respectively, after $10$ tests conducted.}\label{tab3}
\centering
\setlength\tabcolsep{15pt}
\renewcommand{\arraystretch}{1.5}
\begin{tabular}{l cc}
        \hline \hline
         & \textbf{NSFs} (s) & \textbf{Bilby dynesty} (s) \\
\hline
        Average & 0.8 & 256490 \\
        Maximum & 0.8 & 1471691 \\
        Minimum & 0.8 & 60070 \\
        \hline \hline
\end{tabular}
\end{table}

\begin{figure}[!htp]
\centering
\includegraphics[width=0.5\textwidth]{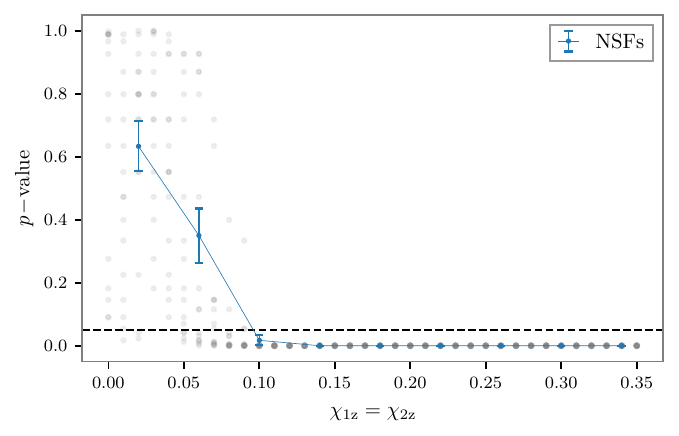}
\centering \caption{\label{fig6}The KS test evaluating the performance of the NSF network on the test set with added spins. 
We assume equal spins for the two black holes ($\chi_{\rm 1z}=\chi_{\rm 2z}$), which increase incrementally from $0$ to $0.35$.
The dot at the center of each error bar represents the mean $p$-value after uniformly binning the 36 datasets. 
The blue line illustrates the trend of these $p$-values across spin values. 
The light gray dots in the background represent $p$-values separately calculated for the 36 spin values, and the color shade indicates the degree of overlap of $p$-values.
The black dotted line indicates a significance level corresponding to a $p$-value of $0.05$.}
\end{figure}

%图6展示了NSF网络在加入不同自旋值的c测试集上的后验推断性能。该图描绘了24个自旋值下p值的误差条分布情况。对于每个自旋值，我们从对应的测试集中随机抽取一个样本（样本大小为2000），并进行十次随机采样（每次采样包含200个事件）以计算十个p值。这些原始的p值在图中以浅灰色圆点表示。为了更清晰地展示网络性能随自旋值的变化趋势，并排除注入噪声的影响，我们将24个自旋值下的p值每4个分为一组进行分箱统计，并绘制了误差条。误差条的中心位置代表该组p值的均值，其长度标记了95%置信区间的范围。结果表明，当自旋值小于约0.32时，p值普遍高于0.05，这表明NSF网络在此范围内具有良好的泛化能力。然而，随着自旋值的增加，p值整体呈现下降趋势；当自旋值大于0.33时，p值显著低于0.05，这可被视为当前训练条件下NSF网络对假设自旋值情形的泛化性边界。
Figure~\ref{fig6} shows the posterior inference performance of the NSF network on the test set that includes different spin parameters.
This figure depicts the distribution of error bars for $p$-values under $36$ spin values.
For each spin value, we randomly select one sample from the corresponding test set (with a sample size of 2000) and perform $16$ random samplings (each sampling containing 200 events) to calculate $16$ $p$-values. 
These original $p$-values are represented by light gray dots in the figure.
To more clearly show the trend of network performance with respect to spin values and to eliminate the influence of injected noise, we divide $p$-values under $36$ spin values into groups of $4$ for binning statistics and draw error bars. 
The center position of the error bars represents the mean of $p$-values in this group, and their length indicates the range of the $95\%$ confidence interval.
The results show that when the spin value is less than approximately $0.1$, $p$-values are generally higher than $0.05$, indicating that the NSF network has good generalization ability within this range.
However, as the spin value increases, $p$-values show an overall downward trend.
When spin value is greater than $0.1$, $p$-values are significantly lower than $0.05$, which can be regarded as the generalization boundary of the NSF network for the assumed spin value scenario under the current training conditions.

%误差棒中间的圆点表示对24组数据进行均匀分箱后p值的平均值，蓝色细线演示了p值的变化趋势。背景中的浅灰色圆点代表24个自旋值下分别统计的p值，颜色深浅代表p值的重合程度。

\section{Conclusion and outlook}\label{sec4}

%1.我们研究了NSF网络对微引力透镜化GW事件进行参数推断方面的适用性。具体而言，我们训练了一个NSF网络来推断点质量透镜模型中描述透镜和源的11维参数。通过KS检验对所有参数的联合检验表明该网络可以给出透镜化事件的合理的参数推断结果。对于随机选取的事件，该网络的推断速度远快于bilby dynesty。虽然精度仍然尚待提高，但在整体准确性方面也与其接近。同时，NSF网络对于训练时未加入的自旋参数也有较好的泛化性，意味着它可能能够适应更多不同类型的事件。
In this work, we investigated the applicability of the NSF network in parameter inference for millilensed GW events. 
Specifically, we trained an NSF network to infer the 11-dimensional parameters describing the lens and source in the point-mass lens model. 
The joint test of all parameters by the KS test indicates that the network can provide reasonable parameter inference results for lensed events.
For randomly selected events, the inference speed of this network is much faster than that of {\tt Bilby dynesty}. Although the accuracy still needs to be improved, it is close to that of {\tt Bilby dynesty} in terms of overall accuracy.
Moreover, the NSF network demonstrates good generalization for spin parameters not included in the training, indicating that it may be capable of adapting to a wider variety of events.

%2总之，本研究证明了NSF能显著提升参数空间内的搜索效率,可以高效完成对微引力透镜化GW事件的参数推断。与传统的贝叶斯方法相比，NSF在13维参数上的平均推断速度比bilby dynesty快四个数量级。这使其有望为透镜化GW事件的低延迟搜索提供强有力的支持。
In conclusion, this study demonstrates that NSFs can significantly enhance the search efficiency within the parameter space and can efficiently complete the parameter inference for millilensed GW events.%更加与摘要呼应
Compared with the traditional Bayesian method, NSFs' average inference speed in an 11-dimensional parameter space is four orders of magnitude faster than that of {\tt Bilby dynesty}. 
This makes it promising to provide strong support for low-latency searches of lensed GW events.

%值得提出的是，尽管我们在本文中仅采用了 PML 模型，但透镜模型中仍存在诸多不确定性。考虑到透镜模型的不确定性，需要搜索更多的额外参数空间。NSFs 网络提供了一种在包含透镜模型的极高维度参数空间下进行参数推断的更为快速的方法。
It is worth noting that although we just adopt PML model in this paper, there exist many uncertainties in lens models. 
Considering the uncertainties on lens models, much more extra parameter space needs to be searched. 
The NSF network provides a much more rapid method for parameter inference under a very high dimension parameter space including lens models.%郭老师增，圆只考虑PML模型之说，学生已收到

%出于简化计算的目的，我们当前的研究采用了简单的GO近似来计算F(f),未来我们可将其拓展至波动光学范围或者采用其他更为精确的计算方式。比如，在文献[]中，通过在低频时使用波动光学，在高频时采用GO光学近似，一定程度上实现了计算效率与精确性的兼顾。但不管怎样，对于透镜化波形数据的模拟过程都是相似的。
What's more, for the purpose of simplifying the calculation, our current research adopts a simple GO approximation to calculate $F(f)$. 
We can expand it to the range of WO or adopt other more precise calculation methods \cite{2020PhRvD.102l4076G,2023PhRvD.108d3527T} in the future. 
For example, in Ref. \cite{Nerin_2025cvae}, by using WO at low frequencies and GO approximation at high frequencies, a certain degree of balance between calculation efficiency and accuracy is achieved. 
However, in any case, the simulation process for lensed waveform data is similar.

%2.未来，我们将进一步探讨NSF网络对于具体参数的推断效果（比如并和时间tc），这可能将涉及到对NSF网络的架构本身的改进。并研究利用NSF网络可快速缩小这些参数的估计范围这一特性，将网络对于这些参数的后验估计作为传统方法的先验，进而完成整个参数推断的过程。以此来探究NSF网络与传统参数推断方法的联合使用。此外，在使用对于大部分透镜化GWs而言，其实同时存在强透镜和微透镜作用，也有研究表明强透镜效应将使得微透镜效应更易检测[ ]，进一步研究这种联合作用下的参数推断效果将是更为实际且有价值的。
In the future, we will further explore the inference effect of the NSF network on specific parameters, which may involve improvements to the architecture of the NSF network itself.
We will also explore the use of the NSF network's ability to quickly narrow down the inference range of these parameters, and use the network's posterior estimates of these parameters as the prior for {\tt Bilby dynesty} to complete the entire parameter inference process \cite{2021PhRvD.103j3006W,Nerin_2025cvae}. 
This will help us explore the combined use of the NSF network and traditional parameter inference methods.

\section*{DATA AVAILABILITY}
The data that support the findings of this article are not publicly available upon publication because it is not technically feasible and/or the cost of preparing, depositing,and hosting the data would be prohibitive within the terms of this research project. The data are available from the authors upon reasonable request.
%All datasets are available from the corresponding authors upon reasonable request.

\section*{SOFTWARE AVAILABILITY}
The code used for simulating GW and data processing techniques is used by {\tt Pycbc}. The code for lensing the simulated GW can be requested from the corresponding authors upon reasonable request. 

\acknowledgments
This research has made use of data or software obtained from the Gravitational Wave Open Science Center (gwosc.org), a service of LIGO Laboratory, the LIGO Scientific Collaboration, the Virgo Collaboration, and KAGRA. 
This work was supported by the National Natural Science Foundation of China (Grants Nos. 12473001, 12575049, 12533001, 11975072, 11875102, 11835009, and 12503001), the National SKA Program of China (Grants Nos. 2022SKA0110200 and 2022SKA0110203), the China Manned Space Program (Grant No. CMS-CSST-2025-A02), and the National 111 Project (Grant No. B16009).
Xiao Guo acknowledges the fellowship of China National Postdoctoral Program for Innovative Talents (Grant No. BX20230104).

\bibliography{paper}

\end{document}